\documentclass[twocolumn,aps,prl,superscriptaddress,longbibliography]{revtex4-2}
\usepackage{amsmath}
\usepackage{amssymb}
\usepackage{graphicx}
\usepackage{dcolumn}
\usepackage{xcolor}
\usepackage[normalem]{ulem}
\usepackage{bm}
\usepackage{braket}
\usepackage[colorlinks,citecolor=blue]{hyperref}
\usepackage{multirow}
\usepackage{makecell}
\usepackage{booktabs}
\usepackage{graphicx} 
\usepackage{physics}
\usepackage{slashed}
\usepackage{appendix}

\newcommand{\lf}{\gamma} % Notation for lattice fermion. 

\newcommand{\SWAP}{\operatorname{SWAP}}

\begin{document}

\title{Uncovering Emergent Spacetime Supersymmetry with Rydberg Atom Arrays}
\author{Chengshu Li}
\thanks{These two authors contributed equally.}
\affiliation{Institute for Advanced Study, Tsinghua University, Beijing, 100084, China}
\author{Shang Liu}
\thanks{These two authors contributed equally.}
\affiliation{Kavli Institute for Theoretical Physics, University of California, Santa Barbara, California 93106, USA}
\author{Hanteng Wang}
\affiliation{Institute for Advanced Study, Tsinghua University, Beijing, 100084, China}
\author{Wenjun Zhang}
\affiliation{Department of Physics and State Key Laboratory of Low Dimensional Quantum Physics, Tsinghua University, Beijing, 100084, China}
\author{Zi-Xiang Li}
\affiliation{Beijing National Laboratory for Condensed Matter Physics and Institute of Physics, Chinese Academy of Sciences, Beijing 100190, China}
\affiliation{University of Chinese Academy of Sciences, Beijing 100049, China}
\author{Hui Zhai}
\affiliation{Institute for Advanced Study, Tsinghua University, Beijing, 100084, China}
\affiliation{Hefei National Laboratory, Hefei 230088, China}
\author{Yingfei Gu}
\email{guyingfei@tsinghua.edu.cn}
\affiliation{Institute for Advanced Study, Tsinghua University, Beijing, 100084, China}
\date{December 23, 2024}

\begin{abstract}

 In the zoo of emergent symmetries in quantum many-body physics, the previously unrealized emergent spacetime supersymmetry (SUSY) is particularly intriguing. Although it was known that spacetime SUSY could emerge at the (1+1)d tricritical Ising transition, an experimental realization is still absent.  
In this work, we propose to realize emergent spacetime SUSY using reconfigurable Rydberg atom arrays featuring two distinct sets of Rydberg excitations, tailored for implementation on dual-species platforms. 
In such systems, the spacetime SUSY manifests itself in the respective correlation functions of a bosonic mode and its fermionic partner. 
However, the correlation function of the fermionic mode inevitably involves a string operator, making direct measurement challenging in the conventional setting.  
Here, we leverage the hybrid analog--digital nature of the Rydberg atom arrays, which allows for the simulation of a physical Hamiltonian and the execution of a digital quantum circuit on the same platform. This hybrid protocol offers a new perspective for uncovering the hidden structure of emergent spacetime SUSY.

\end{abstract}

\maketitle

\textit{Introduction.} Spacetime supersymmetry (SUSY) was proposed in the '70s as a new type of symmetry of spacetime and fundamental particles \cite{Gervais:1971ji,Wess:1974tw}. 
It exchanges bosons and fermions and provides appealing solutions to long-standing problems in particle physics, such as the hierarchy problem~\cite{Dimopoulos:1981zb}. 
However, the experimental evidence of SUSY has not been observed up to the highest energy we can reach in particle accelerators so far. In condensed matter physics, an interesting question is to realize SUSY as an emergent symmetry at low energy in quantum many-body systems -- a fascinating phenomenon where the low-energy states possess larger symmetries than the physical Hamiltonian. This phenomenon often occurs when the system is fine-tuned to a quantum critical point \cite{Sachdev2011}. 

Early in the '80s, it was known that spacetime SUSY could emerge at (1+1) dimensional tricritical Ising transition~\cite{Friedan1985,Zamolodchikov1986}. 
From the Landau--Ginzburg perspective, the conventional second-order Ising transition is described by 
the sign change of the quadratic term in 
a $\phi^4$ theory, where $\phi$ is the order parameter. For a more general free energy functional $\alpha\phi^2+\beta\phi^4+\phi^6$, a first-order transition could occur at a critical line with $\alpha>0$ and $\beta<0$. 
The aforementioned tricritical point refers to the meeting point of the above two transitions, i.e. $\alpha=\beta=0$.

Continuous efforts have been made to realize the tricritical Ising transition~\cite{Blume1966,Capel1966,Alcaraz1985,Grover2014,Rahmani2015,
yao1,
OBrien2018,Li2020,Maffi2024} (or other form of SUSY theories~\cite{SSLee,Yangkun,Yao3,
YaoQED,Li2018}). However, an experimental identification of spacetime SUSY is still missing. 
In addition to the challenges in fine-tuning the model to the tricritical point where the SUSY could emerge, there is another fundamental difficulty in the characterization: the fermionic modes that emerge from the underlying bosonic physical degrees of freedom inevitably involve non-local string operators, whose detection often requires unconventional measurement schemes~\cite{Endres2011,Hilker2017,Chiu2019,Leseleuc2019}. 

Rydberg atom array is a versatile and increasingly powerful platform in both the simulation of quantum phases of matter~\cite{
Bernien2017,
browaeys20,Ebadi2021,
Scholl2021,
Semeghini2021,
Fang2024,
Cheng2024} and quantum information processing~\cite{
RevModPhys.82.2313,
morgado2021quantum,
Bluvstein2022,
thompson23,
endres23,
Bernien23,logical}. 
This platform can operate in (1) analog (simulation) mode where the fundamental degrees of freedom are the pseudo-spin doublets consisting of one of the hyperfine levels in the electronic ground state manifold and one Rydberg state; (2) digital (computation) mode where the data qubit is encoded in two hyperfine levels, with the Rydberg state only used to mediate the entangling gates~\cite{RevModPhys.82.2313}; more interestingly (3) anlog--digital hybrid mode which allows to efficiently generate the desired quantum many-body wave function in analog mode and coherently extract the quantum correlation in the complex wave function with a digital quantum circuit.

In the present work, we propose to overcome the aforementioned challenges in the search for emergent spacetime SUSY using an analog--digital hybrid Rydberg atom array. 
We demonstrate that the flexibility in the interaction type, such as the sign structure of the Rydberg interaction when distinct excitation levels are occupied, 
together with the
geometric reconfigurability, facilitates the realization of a tricritical Ising point in the analog mode. The most important point is that the microscopic parameters, such as detuning and atom separation, can be tuned independently, which allows an unrestricted scan of the entire phase diagram.
This proposal is particularly promising on dual-species platforms~\cite{Zeng2017,Sheng2022,Singh2022,anand2024cng}, where distinct Rydberg excitation levels can be excited for different species without cross-talk. 
Furthermore, we point out that the hybrid nature~\cite{Bluvstein2022} of the platform is ideal for revealing the hidden structure of the emergent spacetime SUSY, 
through measurements of non-local fermionic correlations using quantum circuits.

\begin{figure}[t]
    \centering
    \includegraphics[width=0.8\columnwidth]{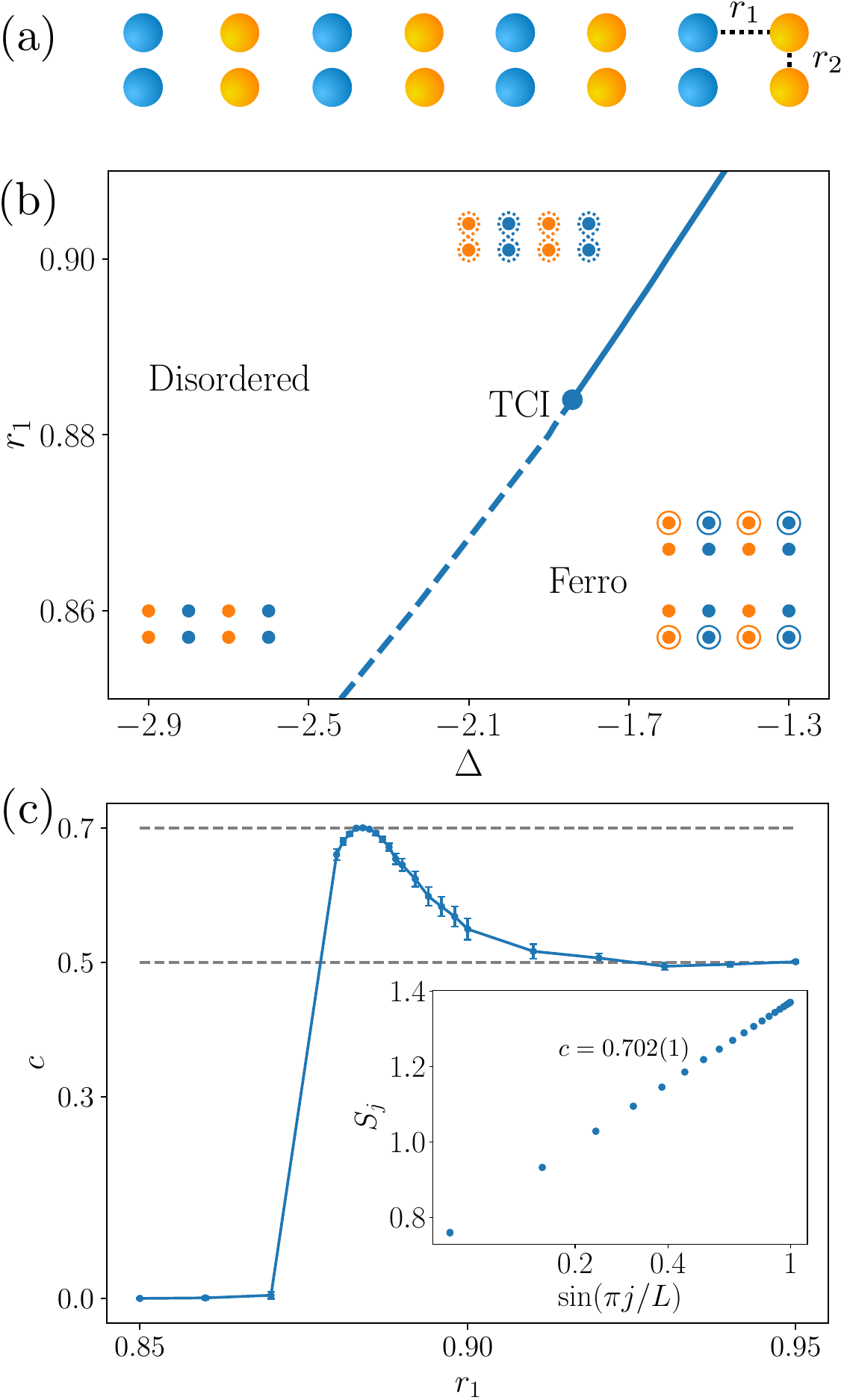}
    \caption{(a) Layout of our model. Two colors denote two sets of atoms subjected to different Rydberg excitation lasers. $r_1$ and $r_2$ denote the nearest separation of atoms along the horizontal and vertical directions. (b) The phase diagram computed for two-leg ladder geometry with $2\times 40$ sites. The solid line denotes a second-order transition, and the dashed line denotes a first-order transition, with ``TCI" denoting the tricritical point. ``Ferro" and ``Disordered" denote the symmetry-breaking phase and the symmetric phase. Here, the $\mathbb{Z}_2$ symmetry is the exchanging symmetry between the upper and the lower legs. $\Delta$ is in unit of $\Omega$ and $r_1$ is in unit of $r_0$ defined in the main text. 
    Dots with a circle denote atoms in the Rydberg state. 
    (c) The central charge $c$ obtained from the entanglement entropy at different given values of $r_1$. $c=7/10$ is the central charge for the tricritical Ising point. The inset shows the fitting of the subsystem entanglement entropy $S(j)$ as a function of the subsystem size $j$ using the Calabrese--Cardy formula. The phase diagram is obtained by the DMRG calculation with bond dimension $200$. The error bars are estimated by fitting the central charge using different ranges of subsystem size and calculating the standard deviation thereof.}
    \label{phase-diagram}
\end{figure}

\textit{Model.} We consider a two-leg ladder geometry of atom arrays as shown in Fig.~\ref{phase-diagram}(a)
where the 
horizontal and vertical separations between atoms are denoted by $r_1$ and $r_2$ respectively. 
The two colors denote the two sets ($A$ and $B$) of atoms subjected to different Rydberg excitation lasers. 
This scheme is readily realized in dual-species atom arrays, where the resonance frequencies of different Rydberg states are largely separated so that crosstalk between different excitations is negligible. 
For simplicity, although not necessary, we assume that two Rydberg excitations share equal coupling strength $\Omega$ and equal detuning $\Delta$. In the following, we set the Rabi frequency $\Omega=1$ as the energy unit. The interaction between two Rydberg atoms is the van der Waals interaction given by $C_{XY}/r^6$ ($X, Y=A$ or $B$). Here, we set $C_{AA}>0$, and we define a length unit $r_0$ by $C_{AA}/r_0^6=\Omega$. 

The Hamiltonian of this two-leg ladder two-sets Rydberg array is given by 
\begin{equation}
\hat{H}=\sum_{i}\left(\frac{\Omega}{2}\hat{\sigma}^x_i-\Delta\hat{n}_i\right)+\sum_{ij}\frac{C_{XY}}{r_{ij}^6}\hat{n}_i\hat{n}_j.  \label{Hamiltonian} 
\end{equation}
It acts on the pseudo-spin states $|\!\!\downarrow \rangle_i$ and $|\!\!\uparrow \rangle_i$ corresponding to the ground state $|g\rangle$ and the Rydberg state $|r\rangle$ of atom $i$. 
$\hat{n}_i=(1+\hat{\sigma}^z_i)/2 $ is the number of Rydberg excitations on site $i$. 
The index $i=(i_x,i_y)$ labels the position of each atom with $i_y=1$ or $2$ denoting the upper or lower leg, respectively. For $i_x$ being odd or even, the atoms belong to $A$ or $B$ subsets. 
Depending on different quantum numbers of Rydberg states, the van der Waals interaction between them can be either repulsive or attractive. 

\textit{Phase Diagram.} 
This two-leg ladder system has a $\mathbb{Z}_2$ symmetry as the exchanging symmetry between the upper and the lower legs. 
Correspondingly, the ground state of this system could exhibit two phases distinguished by whether the $\mathbb{Z}_2$ symmetry is broken, with a first and/or second-order transition between them. If both transitions are present, the meeting point of the two transition lines will be the desired tricritical Ising point we aim for. 

To show that the above conception can be realized in our model with realistic experimental parameters, we set $C_{BB}=1.6C_{AA}$ and $C_{AB}=C_{BA}=-0.97C_{AA}$ as an illustrating example~\footnote{These coefficients are based on the existing dual-species platform, e.g. the Rydberg states of $A$ and $B$ can be the 
${}^{70}S_{1/2}$ and ${}^{73}S_{1/2}$ states of $^{85}$Rb and $^{87}$Rb 
respectively~\cite{ARC}. Similar phase diagrams can be obtained for other choices of atoms, for instance, the Rb--Cs mixture.}
and obtain the phase diagram shown in Fig.~\ref{phase-diagram}(b). 
In the computation, we have also fixed the spacing between the upper and lower legs to be $r_2=0.5 r_0$ so that the repulsive interaction between two Rydberg states with the same $i_x$ is much larger than $\Omega$ and therefore blockades the double occupancy of Rydberg states in the same column. 
The left-over tuning parameters are detuning $\Delta$ and horizontal spacing $r_1$, both are accurately accessible in the reconfigurable atom array platform.

In the phase diagram, the $\mathbb{Z}_2$ symmetry-breaking phase (dubbed ``ferromagnetic'') occurs in the large $\Delta$ and small $r_1$ regime: the larger $\Delta$ induces more Rydberg excitation, while the smaller $r_1$ leads to a stronger attractive interaction between horizontal neighbors (note that $C_{AB}<0$), which favors the configuration that all Rydberg excitations stay in the same row, resulting in the $\mathbb{Z}_2$ symmetry breaking phase. 
Now, starting from the symmetry breaking phase, there are two ways to restore symmetry: (1) shifting toward a more negative value for detuning $\Delta$, which suppresses the Rydberg excitations and ultimately arrives at a symmetric state without Rydberg excitation; (2) shifting towards larger horizontal separation $r_1$, which reduces the attractive interaction and therefore unlocks the correlating effects between different columns. Along this direction, the ground state approximates the product of a symmetric superposition of one Rydberg excitation on the upper row and one on the lower row. It turns out that the phase transition is second order at larger values of $\Delta$ (or, equivalently, $r_1$). Therefore, the meeting point of these two phase boundaries realizes a tricritical Ising transition point on the $(r_1,\Delta)$ parameter space.

Regarding the methodology, the phase diagram Fig.~\ref{phase-diagram}(b) is determined by a combination of the order parameter and the subsystem entanglement entropy, both extracted from the ground state wavefunction numerically calculated 
by the density-matrix renormalization group (DMRG) method~\footnote{
The codes for our numerical calculations are available at https://github.com/chengshul/RydbergSUSY}. 
Here the order parameter we choose is the ``magnetization''
$n^-=n_1-n_2$, where $n_{i_y}=L^{-1}\sum_{i_x=1}^L n_{i}$ with $i=(i_x,i_y)$ and $i_y=1$ or $2$. $n_1-n_2\neq 0$ indicates the $\mathbb{Z}_2$ symmetry breaking. 
The first-order transition is signaled by an abrupt drop in the order parameter. Therefore its location can be precisely determined. In contrast, across the second-order transition, the order parameter vanishes continuously and therefore does not trigger a sharp signal in numerics. 
To precisely determine the location of the second-order phase transition, we compute the central charge by fitting
the entanglement entropy of the subsystem $S(j)$ to the Calabrese--Cardy formula \cite{Calabrese2004} $S(j,L)=\frac{c}{3}\ln\left[\frac{L}{\pi}\sin\left(\frac{\pi j}{L}\right)\right]+S_0$
with $j$ denoting the subsystem size and $L$ the total size. 
A typical example is shown in the inset of Fig.~\ref{phase-diagram}(c). 
The central charge $c$ obtained from the fitting exhibits a maximum at the second-order phase boundary.
Fig.~\ref{phase-diagram}(c) shows the central charge $c$ along the transition line. One can see that $c$ approaches $1/2$ at the Ising transition line, consistent with the prediction of Ising conformal field theory (CFT), and $c$ drops toward zero at the first-order transition. In between, $c$ peaked at the tricritical point with the value $0.702(1)$, consistent with the prediction of supersymmetric CFT.

\begin{figure}[t]
    \centering
    \includegraphics[width=0.9\columnwidth]{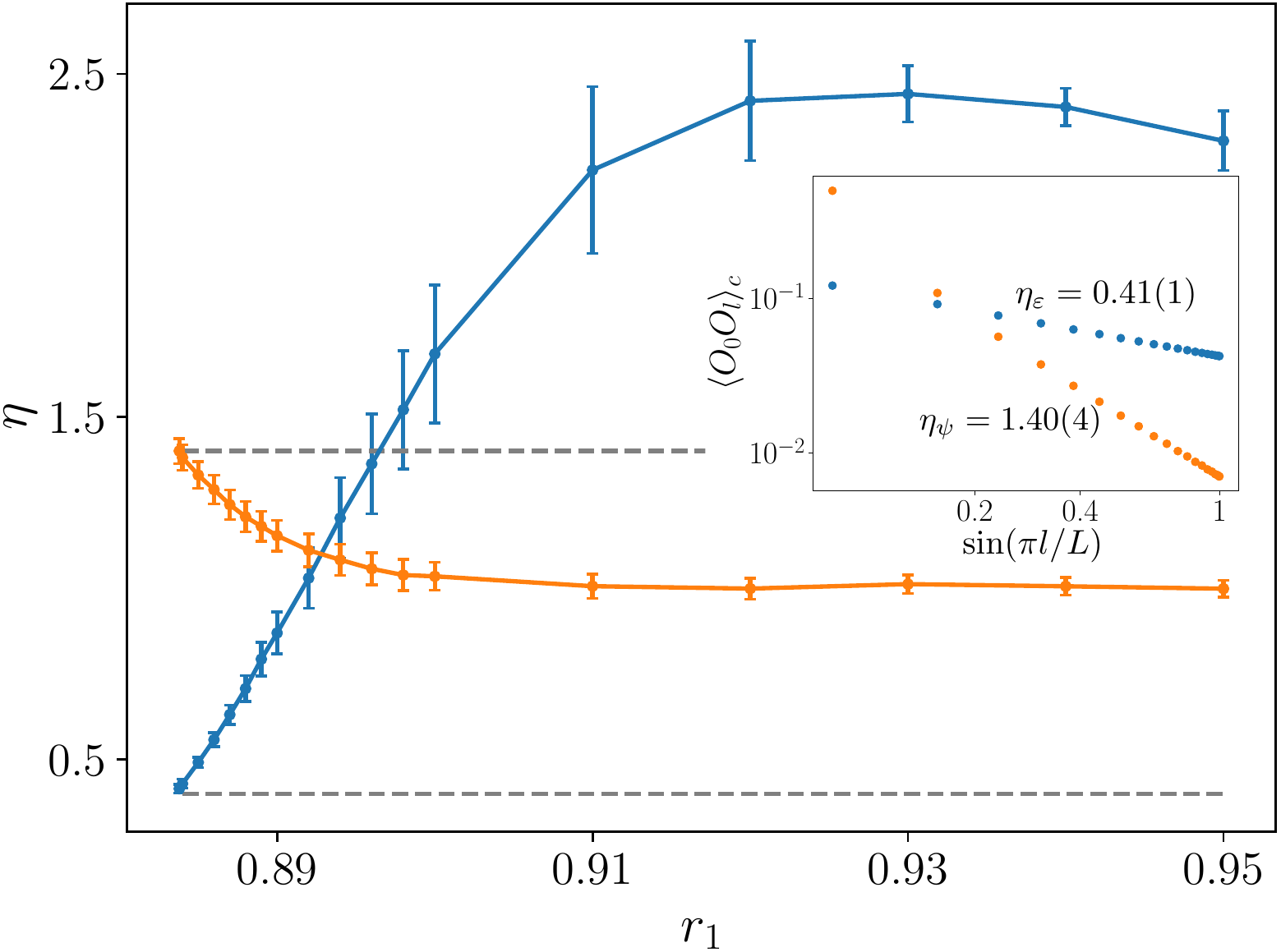}
    \caption{$\eta_\varepsilon$ and $\eta_\psi$ (see text for definition) along the second-order transition line is plotted for different $r_1$.  The inset plots the spatial correlation functions $\langle\varepsilon_{i}\varepsilon_{j}\rangle$ and $ - \left( \langle \gamma'_i \gamma_j\rangle  +\langle \gamma_i \gamma'_j\rangle\right)$ as a function of $|r_i-r_j|$, which is a linear function under the log--log plot. The slope of these two linear functions defines $\eta_\varepsilon$ and $\eta_\psi$. The dashed line indicates that these two numbers differ by unit at the ending point of the second-order transition at $r_1\approx 0.884$. The error bars are estimated by fitting the exponent $\eta$ using different ranges of distance $|r_i-r_j|$ and calculating the standard deviation thereof.}
    \label{SUSY}
\end{figure}

\textit{SUSY Manifestation.} A key physical consequence of the emergent SUSY at the critical point is that the scaling dimensions of a low-lying bosonic mode and its fermionic superpartner differ by $1/2$. 
For a motivating and simplest example, let us consider a $D$ dimensional (spacetime dimension) supersymmetric theory of a free massless scalar paired with a massless free fermion, with Lagrangian $\mathcal{L}_{\rm SUSY}=\mathcal{L}_B[\phi] + \mathcal{L}_F[\psi]$. From the bosonic part $\mathcal{L}_B=\int d^D{\bf r}(\partial\phi)^2$, we get the dimension of $[\phi]$ from equation $2[\phi]=D-2$. Therefore, by dimensional analysis $\langle\phi({\bf r}_i)\phi({\bf r}_j)\rangle\sim {1}/{|{\bf r}_i-{\bf r}_j|^{D-2}}$.
In parallel, the fermionic partner $\psi$ with $\mathcal{L}_F=\int d^D{\bf r}\bar{\psi}i\slashed{\partial}\psi$ satisfies $2[\psi]=D-1$ and, consequently, $\langle\psi({\bf r}_i)\psi({\bf r}_j)\rangle\sim {1}/{|{\bf r}_i-{\bf r}_j|^{D-1}}$. 
The point is that the exponents of these two decayed spatial correction functions of power law differ by $1$, which is a measurable consequence of the SUSY we will investigate in the following. 

In contrast to the aforementioned free theory example, where SUSY is inherent, our microscopic model~\eqref{Hamiltonian} is constructed based on pseudo-spin degrees of freedom, with SUSY emerging only at the tricritical point. In such a scenario, it is generally challenging to identify the SUSY pair of fields that result from a non-trivial renormalization group flow~\cite{Zou2020}. In this work, we exploit the aforementioned $\mathbb{Z}_2$ symmetry to identify the SUSY pair in our model. 
In the field-theoretic description of the tricritical Ising transition, there is a $\mathbb{Z}_2$-even real bosonic field $\varepsilon$ representing the ``energy'', and a natural candidate in our microscopic model is the sum of Rydberg excitation numbers on each rung, by an abuse of notation let us denote $ \varepsilon_{i}=\hat{n}^+_i= \hat n_{(i,1)}+\hat n_{(i,2)}$. Under SUSY, this $\mathbb{Z}_2$-even operator should be related to a pair of $\mathbb{Z}_2$-odd real fermion operators.  
A simple $\mathbb{Z}_2$-odd lattice operator is $\hat n^-_{i}=\hat n_{(i,1)}-\hat n_{(i,2)}$. 
However, $\hat n^-_{i}$ itself is not yet a fermionic operator. To finish the construction, we need to attach a Jordan--Wigner string so that distant fermion operators anticommute. This gives two natural candidates: 
\begin{equation}
    \hat\lf_i=\bigg( \prod\limits_{j=1}^{i-1}\SWAP_{j} \bigg) \hat{n}^-_{i}, \quad
    \hat\lf'_i=\bigg( \prod\limits_{j=1}^{i}\SWAP_{j} \bigg) \hat{n}^-_{i}, 
\end{equation}
where the swap operator $\SWAP_j$ applies an exchange between the upper and the lower legs at site-$j$ onto the quantum state. 
Their linear combination corresponds to the time reversal pair of fermion fields in the continuum field theory, i.e. $\psi= \hat\gamma-i \hat\gamma'$ and $\bar{\psi} = \hat\gamma + i \hat\gamma'$ with abuse of notation.

Now, to establish the SUSY relation, we need to check the exponents of the decayed spatial correlation function of the bosonic $\varepsilon$ and fermionic $\psi$ (or $\bar{\psi}$, as they have the same scaling dimension). For the former, we need to measure the correlation function of the local operator $\varepsilon_{i}=\hat n_{(i,1)}+\hat n_{(i,2)}$, that is,
\begin{equation}
    \langle\varepsilon_{i}\varepsilon_{j}\rangle \sim {1}/{|{\bf r}_i-{\bf r}_j|^{\eta_\varepsilon}},
\end{equation}
and extract the exponent $\eta_\varepsilon$. For the fermionic operator $\psi$, we need to extract the exponent $\eta_\psi$ from the following correlation function
\begin{equation}
    \langle \psi_i \psi_j \rangle = -i \left( \langle \gamma'_i \gamma_j\rangle  + 
    \langle \gamma_i \gamma'_j\rangle
    \right) \sim {1}/{|{\bf r}_i-{\bf r}_j|^{\eta_\psi}}, 
    \label{fermion}
\end{equation}
where in the first step we have used the property that $\langle \gamma_i \gamma_j \rangle=\langle \gamma'_i\gamma'_j\rangle=0$ enforced by the time-reversal symmetry. Note the correlation function $\langle\lf'_{i}\lf_{j}\rangle$ or $\langle \gamma_i \gamma'_j\rangle$
 involves a non-local string operator, for instance
\begin{equation}
    -\langle\lf'_{i}\lf_{j}\rangle=\langle \hat{n}^-_i  \bigg(\prod\limits_{k=i+1}^{j-1}\SWAP_{k}\bigg)\hat{n}^-_{j}\rangle 
    \label{nonlocal}
\end{equation}
The SUSY in our model manifests itself as $\eta_\psi-\eta_\varepsilon=1$.

We show in Fig.~\ref{SUSY} the numerical results for the equal-time correlation functions $ \langle\varepsilon_{i}\varepsilon_{j}\rangle$ and $ - \left( \langle \gamma'_i \gamma_j\rangle  +\langle \gamma_i \gamma'_j\rangle\right)$. The inset confirms the power-law behavior of the two correlation functions at the tricritical point.  
From the slope of the log--log plot, we can extract the exponent as shown in Fig.~\ref{SUSY}. 
It verifies clearly that at the end point of the second-order transition line, these two powers $\eta_\varepsilon$ and $\eta_\psi$ differ by $1$, which is direct evidence of the emergent SUSY. Away from the tricritical point, the relation fails.

\textit{Measuring fermionic correlation on Rydberg atom array.} 
Measuring the non-local string operator in \eqref{nonlocal} with a set of SWAP is generally difficult. The situation is similar to the measurement of the second Renyi entropy where a many-body interference between two identical copies of quantum states is performed~\cite{Zoller12,Greiner15}. 
Here, we need to interfere with two subsets of our single quantum state. 
To accomplish the measurement, 
we can utilize the hybrid nature of the Rydberg atom array platform to perform a quantum version of the ``analog-to-digital converter'' and then process the quantum data via a measurement circuit consisting of entangling gates between the upper and lower chains. 

\begin{figure}[t]
    \centering
    \includegraphics[width=0.8\columnwidth]{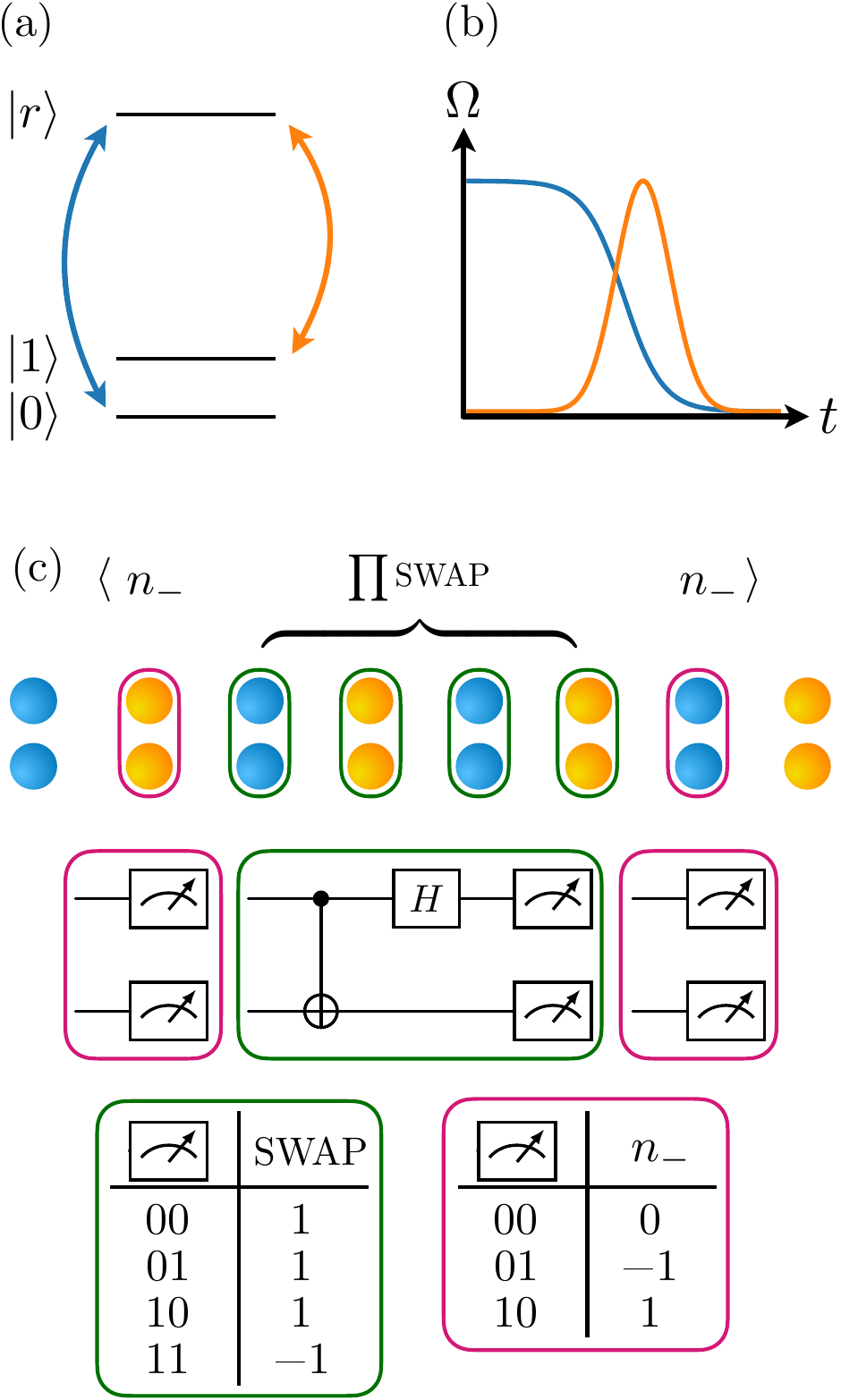}
    \caption{(a) $\ket{0}$ and $\ket{1}$ are two hyperfine levels of the electronic ground state, and $\ket{r}$ is a Rydberg state. $\ket{0}=|g\rangle$ and $\ket{r}$ are used for quantum simulation and $\ket{0}$ and $\ket{1}$ form a basis for digital quantum computation. The blue line denotes the Rydberg coupling $\Omega$ in the many-body Hamiltonian \eqref{Hamiltonian}, and the orange line represents the qADC coupling that transfers a many-body wave function into the quantum computation basis. (b) Schematic of the time sequence of turning off the Rydberg coupling and the AtoD pulse. (c) After the qADC, we use the quantum circuit shown in the box to process the quantum many-body wave function to measure the fermionic correlation function \eqref{fermion}. The tables in the last row summarize the correspondence between the measurement outcome and the eigenvalue of $\SWAP$ and $n^-$.   }
    \label{AtoD}
\end{figure}

A quantum Analog-to-Digital Converter (qADC) and its partner quantum Digitial-to-Analog Converter (qDAC) map between the ``analog basis'' $\{ |g\rangle, |r\rangle \}$ (``ground state'' and ``Rydberg state'') and the ``digital basis'' $\{ |0\rangle, |1\rangle \}$. In the Rydberg atom, these two Hilbert spaces are embedded in the rich atomic levels. For instance, in the $^{87}$Rb atom, the digital basis can be chosen as the two hyperfine spin levels of the ground state, and one of them, say $|0\rangle$ can be used for the ground state $|g\rangle$ in the analog basis as well. For this three-level system, the qADC can be realized by stimulated Raman adiabatic passage (STIRAP) \cite{Zhai2021,STIRAP} where
an additional laser is applied to coherently transfer the Rydberg state $|r\rangle$ to the hyperfine $|1\rangle$ state, while keeping the $|g\rangle = |0\rangle$ state invariant. In this protocol, an optimization between the qADC pulse and the Rydberg laser in the analog mode is expected to mitigate the dephasing error caused by the turning off of the Rydberg laser that couples the analog basis (See End Matter for detailed discussion).
Alternatively, one can also realize the qADC via a two-step process, named the coherent mapping protocol in Ref.~\cite{Bluvstein2022}, where the ground state $|g\rangle$ (identified as $|1\rangle$ in this protocol) is transferred to digital $|0\rangle$ and then $|r\rangle$ is transferred to $|1\rangle$.

Following the qADC, we can utilize the quantum computation architecture to process the many-body wavefunction in the digital basis where $|0\rangle$ represents the empty states and $|1\rangle$ represents the states with Rydberg excitation. For each rung in the ladder, we have three allowed states $|00\rangle$, $\frac{1}{\sqrt{2}} (|01\rangle+|10\rangle)$, $\frac{1}{\sqrt{2}}(|01\rangle-|10\rangle)$ with eigenvalue $1$,$1$,$-1$ under SWAP respectively. To single out the last state, we run a measurement circuit shown in Fig.~\ref{AtoD}(c) and collect the instance with measurement outcome $|11\rangle$ in the computation basis. Then the fermionic correlator \eqref{fermion} is obtained as 
\begin{equation}
    \langle (-1)^{{\rm \# |11\rangle~ pairs~between}~(i,j)} ~ \hat{n}^-_i \hat{n}^-_j \rangle.
\end{equation}
Namely, the measurement of $\langle \psi_i \psi_j \rangle$ amounts to averaging the measurements of two local operators $ \hat{n}^-_i$ $\hat{n}^-_j $, with an additional phase $-1$ assigned to all the instances with odd number of $|11\rangle$ between site $i$ and $j$. We note that measuring the string of swap operators only requires uniformly applying the quantum circuit shown in Fig.~\ref{AtoD} to all sites where swap operators are located, which can be achieved by first moving these qubits to the operation zone using the coherent transport technique~\cite{Bluvstein2022} and applying these gates in parallel there.

\textit{Summary and Outlook.} In this work, we show that a two-leg ladder Rydberg array structure with two sets of Rydberg excitations is sufficient to realize the tricritical Ising transition where the spacetime SUSY emerges. This proposed setup is especially suitable for the dual-species Rydberg atom platforms where the two Rydberg excitations can be implemented without cross-talk. 
We would like to remark that unlike the single chain model discussed in literature \cite{Fendley2004,Bernien2017,Slagle2021} where the Ising symmetry is realized as a translation by one lattice spacing, here the Ising $\mathbb{Z}_2$ symmetry is realized as the exchanging between two rails of the ladder. Combining this geometric realization of the internal symmetry with the reconfigurability of the atom array, 
one can further study how various boundary conditions can be realized geometrically. We would like to mention that a quick diagnosis of the tricritical point without resorting to correlation measurements can be carried out with a Kibble--Zurek ramping~\cite{Kibble1976,Zurek1985,Keesling2019}, with the universal exponent $\mu=5/14$. In this work, we show that a two-leg ladder Rydberg array structure with two sets of Rydberg excitations is sufficient to realize the tricritical Ising transition where the spacetime SUSY emerges, which also represents a qubitization of the supersymmetric field theory~\cite{Bhattacharya2021,Caspar2022,Malti2024}.

To manifest the emergent SUSY encoded in the many-body wavefunction, we leverage the analog--digital hybrid nature of the Rydberg atom platform.  This involves directly loading the ``experimental data'' prepared by the analog mode to the ``quantum computer'' in the digital mode via a quantum Analog-to-Digital converter (qADC). By processing the data through quantum circuits, we can access the information that is difficult to retrieve in conventional experiments where the data are processed classically~\cite{QAM22,QAL22,lu2024digital}. 

Additionally, we envision that a quantum Digital-to-Analog Converter (qDAC) holds potential in the quantum simulation of many-body physics. Using the quantum circuit module, more interesting initial states can be prepared and sent to the quantum simulator, opening new possibilities for exploring many-body quantum dynamics.

\textit{Acknowledgments}. 
We thank  
Wenlan Chen, Xun Gao, Jiazhong Hu, Xiaodong He, Xiao-Liang Qi, Kunpeng Wang, Peng Xu and Tao Zhang for discussions. 
We acknowledge financial support from 
the Innovation Program for Quantum Science and Technology 2021ZD0302005, 
the NSFC Grant No.~12042505 and U23A6004, 
the National Key R\&D Program of China 2023YFA1406702, 
and the Tsinghua University Dushi Program. 
S.L. is supported by the Gordon and Betty Moore Foundation under Grant No.\,GBMF8690 and the National Science Foundation under Grant No.\,NSF~PHY-1748958.
Z.X.L. is supported by the NSFC under Grant No.~12347107.
H.Z. acknowledges support from the XPLORER Prize. 
Y.G. acknowledges support from the DAMO Academy Young Fellow program. 
The DMRG calculations are performed using the ITensor library (v0.3) \cite{ITensor}.

\appendix
\section{End Matter}
\subsection{STIRAP time sequence}
\begin{figure}[t]
    \centering
    \includegraphics[width=0.95\linewidth]{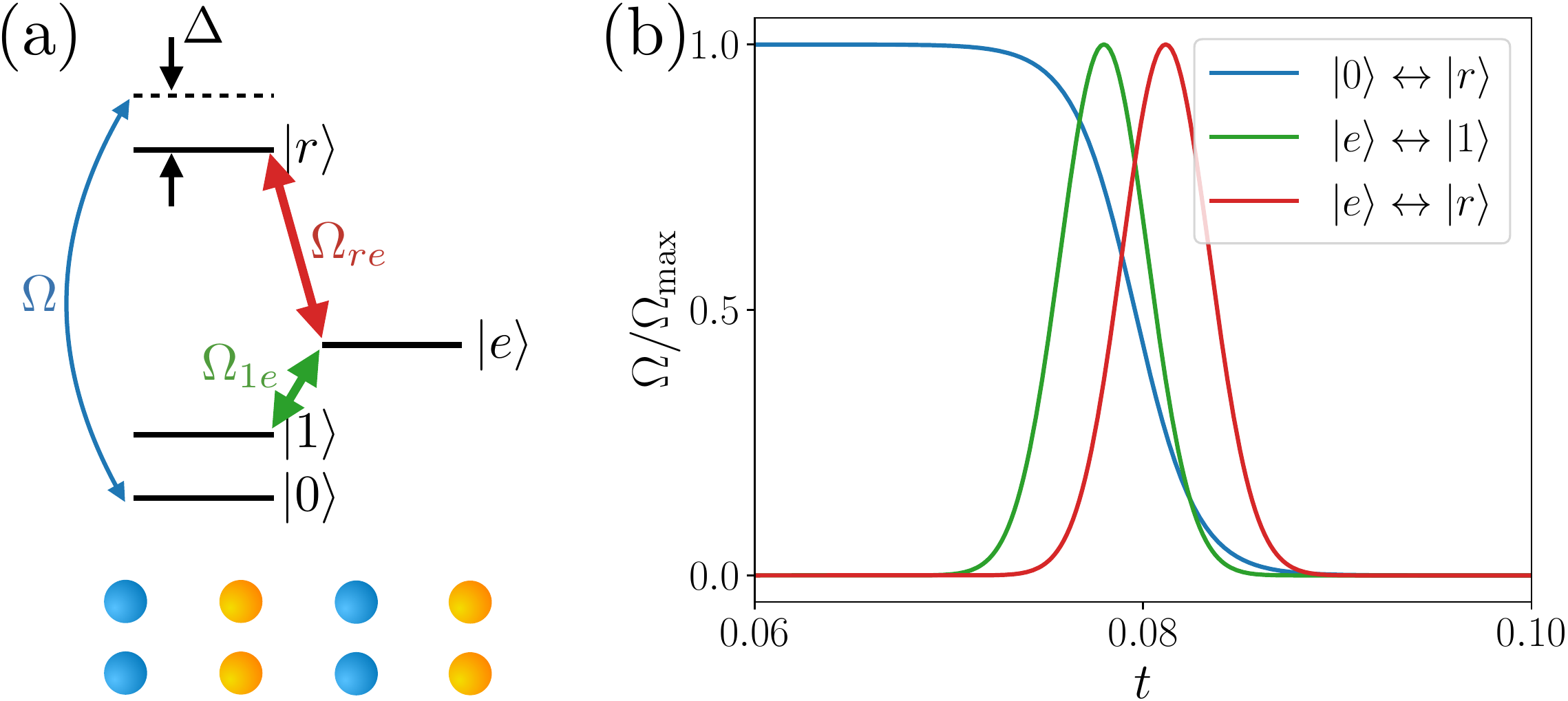}
    \caption{(a) Schematic of the laser coupling between the ground state hyperfine levels $\ket{0}$ and the Rydberg level $\ket{r}$ (blue line, denoted by $\Omega$), the coupling between another ground state hyperfine level $\ket{1}$ and the intermediate state $\ket{e}$ (green line, denoted by $\Omega_{1e}$), and the coupling between $\ket{r}$) and $\ket{e}$ (red line, denoted by $\Omega_{re}$). (b) The time sequence of these three laser couplings. Here, the couplings of laser strengths $\Omega$ are normalized by their corresponding maximum value $\Omega_\text{max}$. The bare detuning for $\Omega_{1e}$ and $\Omega_{re}$ are set as zero.  }
    \label{fig:AtoDschematic}
\end{figure}

This section presents a numerical simulation of the quantum dynamics that enables Analog-to-Digital conversion. This dynamical process involves gradually turning off the Rydberg coupling light that couples $\ket{0}$ and $\ket{r}$ and simultaneously turning on a stimulated Raman adiabatic passage (STIRAP) process \cite{Zhai2021,STIRAP} to transfer atoms from $\ket{r}$ state to $\ket{1}$ state, as shown in Fig.~\ref{fig:AtoDschematic}(a). For each atom, the calculation takes into account four internal states, namely $\ket{0,1,r,e}$, where $\ket{e}$ is an intermediate state introduced for STIRAP. The control of the time sequence of three lasers is shown in Fig.~\ref{fig:AtoDschematic}(b). First, the laser coupling between $|e\rangle$ and $|1\rangle$ are turned on and off, and subsequently, the laser coupling between $\ket{e}$ and $\ket{r}$ are turned on and off, which completes the STIRAP process. In the meantime, the laser coupling between $\ket{r}$ and $\ket{0}$ is gradually turned off. 

\begin{figure*}
    \centering
    \includegraphics[width=0.95\linewidth]{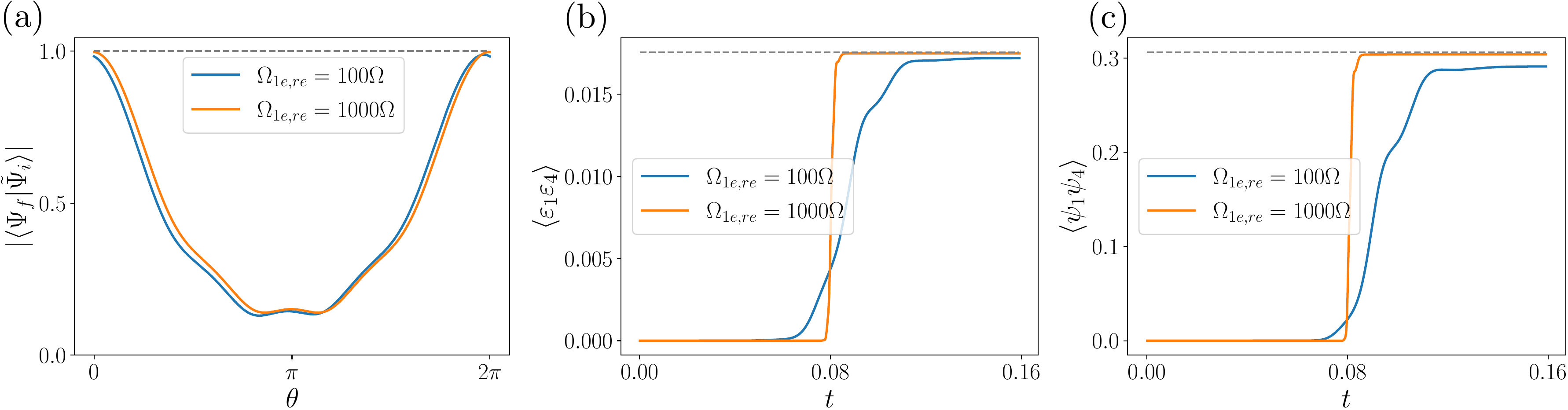}
    \caption{(a) The wave function overlap between $\Psi_f$ and $\tilde{\Psi}_i$ as a function of $\theta$. (b, c) Bosonic $\langle\epsilon_1\epsilon_4\rangle$ and fermionic $\langle\psi_1\psi_4\rangle$ correlators between the first and the fourth rung under $\Psi_f$ as a function of the time evolution. The dashed lines indicate the initial values of the same correlations under $\Psi_i$. This calculation is done with $r_1=0.88$, $r_2=0.5$ and $\Delta=-1.3$. In all calculations we set the maximum value of the STIRAP Rabi frequency $\Omega_{1e}=\Omega_{re}$ as  $100\Omega$ and $1000\Omega$ respectively, where $\Omega$ is the Rydberg coupling strength defined in the main text.}
    \label{fig:AtoDeffect}
\end{figure*}

We numerically calculate this dynamical process for a system with a total of four rungs (eight sites). We start with the ground state of this system denoted by $\Psi_i$ and numerically simulate the dynamical evolution process following the time sequence shown in Fig.~\ref{fig:AtoDschematic}(b). This dynamical evolution leads to a final state denoted by $\Psi_f$. To quantify $\Psi_f$, we note that $\Psi_i$ is initially written in $\{\ket{0},\ket{r}\}$ bases, and if we assume perfect Analog-to-Digital conversion, we can perform a basis transformation by replacing $\ket{r}$ with $e^{i\theta}\ket{1}$ in $\Psi_i$, which leads to $\tilde{\Psi}_i$. 
In Fig.~\ref{fig:AtoDeffect}(a), we show the wave function overlap between $\Psi_f$ and $\tilde{\Psi}_i$ as a function of $\theta$. This overlap can reach a maximum value very close to one for an optimal value of $\theta$, as shown in Fig.~\ref{fig:AtoDeffect}(a). This shows that this dynamical process can achieve a perfect Analogy-to-Digital conversion. Since $\theta$ is a gauge choice in $\tilde{\Psi}_i$, all physical observables should not depend on $\theta$. Furthermore, given the nearly perfect overlap between $\tilde{\Psi}_i$ and $\Psi_f$, the correlation functions under $\Psi_f$ should uncover the corresponding correlations under $\Psi_i$.
To confirm this, we compute the bosonic and fermionic correlators under $\Psi_f$, where we should also modify the definition of $\epsilon$ and $\psi$ operators accordingly by replacing the role of Rydberg state $\ket{r}$ with $\ket{1}$ state. Two typical examples of the results are shown in Fig.~\ref{fig:AtoDeffect}(b) and (c). We should compare the result after the dynamic process with the initial correlation under $\Psi_i$, as shown by the dashed lines, and we find that perfect agreements between these two are reached.

\bibliography{biblio}

\end{document}